# Evaluating Automated Testing on an Open-Source Web Application Using Cypress

Quoc-Binh Nguyen
*Faculty of Information Technology*
*Ton Duc Thang University*
Ho Chi Minh City, Viet Nam
nguyenquocbinh@tdtu.edu.vn

Truc-Ly Phan Nguyen
*Computer Science Program*
*Vietnamese-German University*
Ho Chi Minh City, Viet Nam
18715@student.vgu.edu.vn

Ngoc Hong Tran
*Computer Science Program*
*Vietnamese-German University*
Ho Chi Minh City, Viet Nam
ngoc.th@vgu.edu.vn

Dung Hai Dinh*
*Business Information System Program*
*Vietnamese-German University*
Ho Chi Minh City, Viet Nam
*Corresponding author: dung.dh@vgu.edu.vn

***Abstract*—End-to-end automated testing is increasingly used in web software development to ensure system quality and shorten response times during development. However, the true effectiveness of automated testing depends on many factors including execution time, stability of test results, and maintainability of the test suite as the application continues to evolve. In this paper, we evaluate the effectiveness of end-to-end automated testing using the Cypress framework for an open-source web application. We deployed the experiment with 27 test cases. The test's effectiveness is measured by execution speed, reliability, and maintainability. The experimental results show that the Cypress-based end-to-end test suite has short and stable execution times. It suits frequent runs during software development. The majority of test cases achieved consistent results across multiple runs, while flakiness only occurred in a few tests which involve complex interactive functions. Furthermore, the study highlights the impact of element locator strategies and Page Object Model (POM) architecture on test suite maintainability, demonstrating that resilient data-cy attributes significantly reduce maintenance overhead when UI changes occur.**

## I. Introduction

In the modern digital era, the software quality is not an option but an essential part for both business corporations and economic stability. However, maintaining this quality faces unprecedented challenges in terms of expense and technique. Research from the National Institute of Standards and Technology (NIST) pointed out that inadequate infrastructure of software testing caused damage to the US economy at approximately 59.5 billion dollars annually **[1]**. This number has reached trillions of dollars due to the increasing complexity of software. The fundamental reason is the "error cost curve". If an error is identified during the operating period, it can cost up to 100 times more than if it were detected early on **[2]**. The transition to accelerated development frameworks such as Agile **[3]** and DevOps **[4]** has exerted significant strain on the quality assurance (QA) process. Manual testing has become a constraint due to its inability to keep pace with the constant deployment of new features **[5]**. In open-source contexts, the contributions are made by volunteers without a dedicated QA department, so the absence of an efficient automated testing mechanism frequently results in a deterioration of software quality over time **[6]**.

Under above motivation, in this paper, we evaluated the effectiveness of Cypress through a case study on an open-source web application. The goal is to examine whether this tool can realistically address the long-standing trade-off between development speed and software reliability. Although automated testing is often presented as a practical solution, its actual benefits remain uncertain in many real-world projects. In practice, automated test suites are frequently affected by issues such as flakiness and high maintenance overhead, especially when user interface (UI) changes occur **[19]**. These challenges can reduce their reliability and limit their usefulness over time. To investigate this further, we apply Cypress to the open-source real world project and conduct an empirical evaluation of its performance. Beyond replacing manual testing efforts, this work explicitly evaluates how different element locating strategies (traditional vs. data-cy attributes) combined with a Page Object Model (POM) architecture impact the long-term maintainability of the test suite.

This paper is organized into 6 sections. The problem statement and contribution are presented in Section I. The fundamental software testing is presented in Section II. Section III presents the automated testing. The experimentation is stated in Section IV, and the result findings are described in Section V. Eventually, the conclusion and future works are presented in Section VI.

## II. Software Testing

Software testing has been an activity intertwined with the history and development of the digital computing industry. It is considered an essential method for evaluating and determining the quality of a software system before it goes into operation. In reality, testing activities typically account for 40% to 50% of total project development resources, and this figure increases for systems requiring high stability and reliability **[7]**. According to E. Dustin et al. **[8]**, investing significant resources in testing is a strategic necessity to control risk and ensure product sustainability throughout its development cycle. Software testing contains a series of activities, which is called the Software Testing Life Cycle (STLC) presented in Fig. 1.

### *A. A typical STLC process*

Including 6 following stages:

*1) Requirement analysis:* The testing team studies the requirements documents to identify testable features and non-functional requirements.

*2) Test Planning:* This is the most important stage, determining the testing strategy, selecting tools (such as Cypress), and estimating resources and risks.

*3) Test Case Development:* Engineers create detailed test scenarios, test data, and automation scripts.

*4)* Test Environment Setup: Hardware, software, and network conditions are prepared to ensure the test environment closely resembles the user's real-world environment.

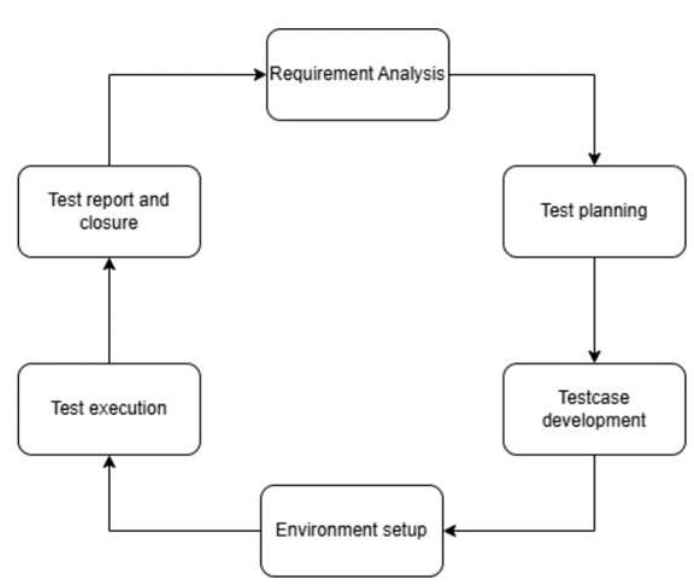

Fig. 1. Software Development Life Cycle

*5) Test Execution:* Tests are run, and bugs are recorded for the development team to fix. In the CI/CD model, this phase is often fully automated to provide rapid feedback.

*6) Test Cycle Closure:* The team summarizes the results, reports on test coverage, and draws lessons learned to improve the process for the future.

A well-structured implementation of STLC closely connects testing activities with the overall Software Development Cycle (SDLC). As a result, the software delivery process gets more stable and less risky.

### B. Testing in Modern Software Development

*1) Waterfall.* is the most well-known and traditional method of developing software, in which the software processes are separated into different phases, and each phase follows each other in an irreversible straight line **[9]**. In this model (see Fig. 2), the testing phase is placed at the end of the development lifecycle as the terminal gatekeeper. However, this model contains a systematic flaw in terms of risk management. Since the approach is sequential, each phase in the model is "signed off", meaning the next phase can only be started after the previous one is completed. This can be a significant drawback when unforeseen circumstances and changes arise during the project **[10]**.

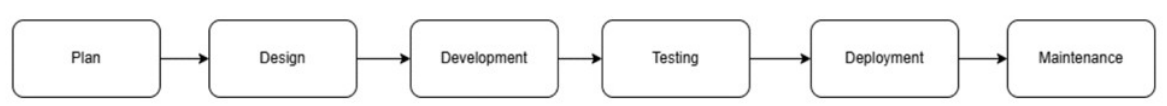

Fig. 2. Waterfall model

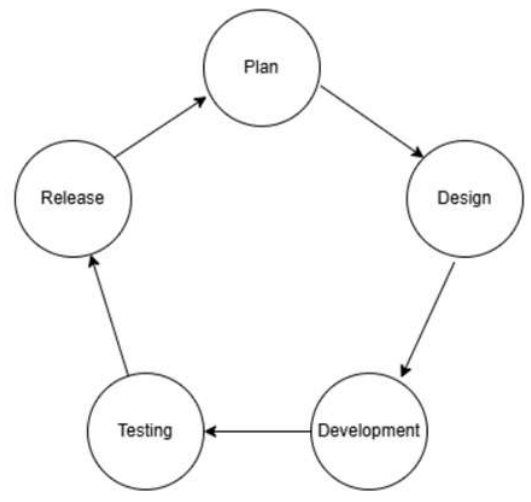

Fig. 3. Agile Methodology

*2) Agile.* emerges as an ideal solution to the limitations of Waterfall, emphasizing flexibility and collaboration in order to meet client expectations and adapt to changing market circumstances **[10]** (see Fig. 3). In this model, testing is no longer a phase but a continuous activity throughout the development lifecycle **[11]**. By breaking the development down to short iterations, Agile encourages constant feedback loops, allowing teams to discover and address issues early on, hence lowering the expense and effort needed to address problems later in the lifecycle **[10]**. Moreover, since testing tasks appear in every sprint, automated testing becomes an ultimate solution against regression.

## III. Automated Testing

While manual testing relies on human observation and interactions, Automated testing is the technique of controlling test execution and comparing actual results with expected results using software that is distinct from the software being tested **[8]**. There are three levels of automated testing (see Fig. 4). At the bottom, unit test concentrates on testing a singular component in the application, such as separated methods and code functions, thus it provides specific data and errors with high precision. At the middle level, service test checks the interactions between components and makes sure they work together without any problems. At the top of the pyramid, End-to-End test (E2E) simulates the end-user experience across the whole application. It ensures that different layers of the system, from frontend to backend, database, and integrations, work as intended. E2E can be performed both manually and automatically.

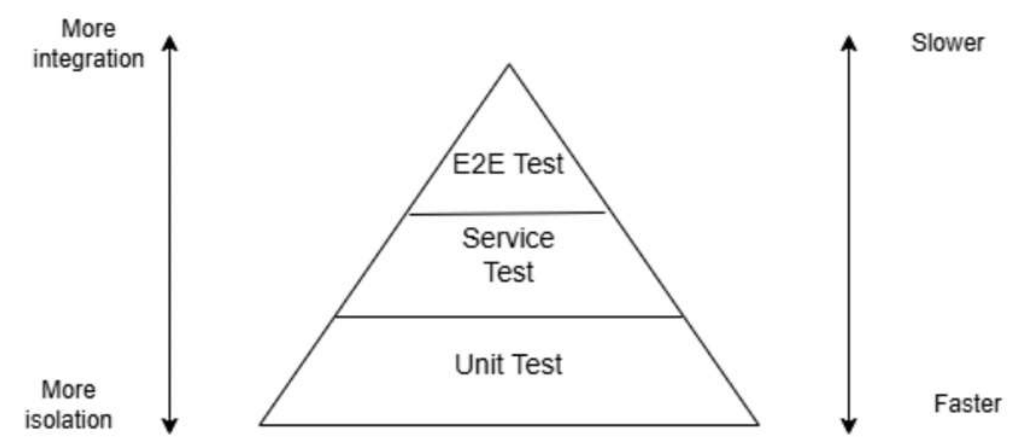

Fig. 4. Testing pyramid **[12]**

In addition, automated testing plays a strategic role as a pillar in the quality assurance process, especially for projects operating under Agile and DevOps models. Test automation is a great candidate for regression testing. "Any tedious or repetitive task involved in developing software is a candidate for automation" **[13]**. As the application gets bigger after every short iteration, the time of the test also grows longer, sometimes exponentially depending on the application under test (AUT). Furthermore, automated testing is an integral component of Continuous Integration and Continuous Delivery (CI/CD) **[14]**. Integrating tests into the deployment pipeline provides fast feedback, allowing the team to detect and address errors as soon as they arise. Running a full test suite automatically saves the team a lot of time, produces more accurate results, and frees people from exhausting tasks **[13]**.

### A. Cypress framework

Cypress is a Javascript based testing framework introduced in 2015 by Cypress.io under the MIT license **[15]**. Designed specifically as a frontend testing tool for modern web applications, Cypress provides solutions for E2E testing, component testing, UI testing, and many more. Cypress products include Cypress app, an open-source app for writing and running tests locally, Cypress Cloud which is a paid service for recording test results and providing test analytics, Cypress Accessibility as a premium solution for accessibility checks, and UI Coverage for visual overview of test coverage across web pages and components.

### B. *Cypress vs Selenium*

While traditional testing tools like Selenium rely on communication between Web driver protocol and Document Object Model (DOM) across HTTP request, Cypress works directly within the browser environment by utilizes a universal driver that is compatible with modern web browsers like Firefox, Edge, Chrome and other Chromium browsers **[16]**. By operating test inside browser runtime, Cypress allows users to interact with application under test in real time.

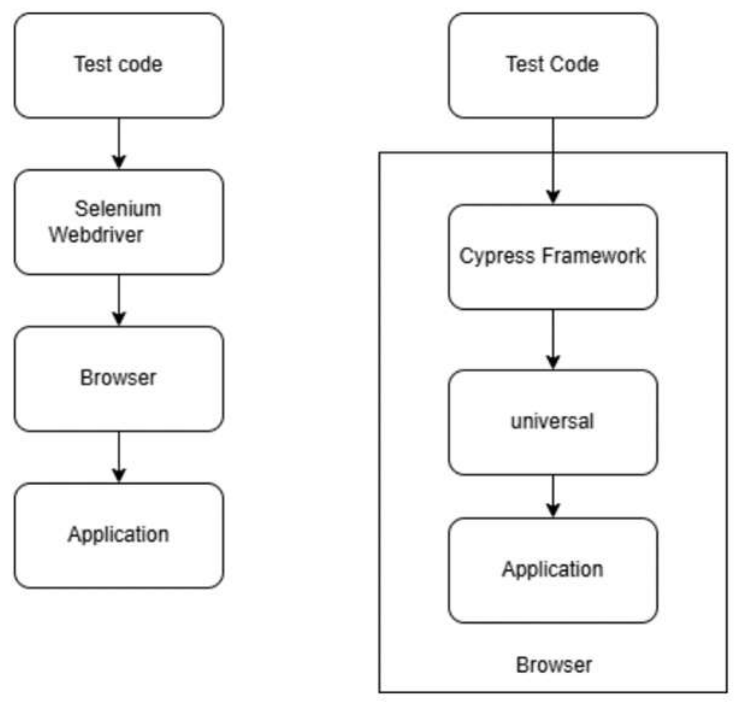

Fig. 5. Comparison between Selenium and Cypress Framework

This architecture also allows Cypress to reduce timing issue, which often arise in traditional testing tool because the commands must travel through multiple layers. This architecture also has an inherent advantage because it incorporates automatic waiting methods that synchronize test execution with application events, including network requests or DOM updates. On the other hand, in order to handle asynchronous action, Selenium-based tests frequently call for explicit or implicit waits. Additionally, using a universal driver guarantees that tests run consistently in various browsers.

Regardless of the browser in which they are run, Cypress test cases adhere to a consistent format and behavior. By using the same test suite on newly supported browsers, QA engineers and developers can increase cross browser testing efforts without having to make major modifications to the current tests. The below figure shows the difference in architecture between Selenium and Cypress. While test execution takes place outside the browser in Selenium, Cypress test is executed inside the browser **[16]**. Although Cypress supports only test development in JavaScript and Typescript, which limits the language flexibility compared to Selenium, it aligns well with contemporary frontend development techniques.

## IV. EXPERIMENTATION

### A. *Set-up Environment*

We deploy the experiment using a local machine equipped with Intel Core i7 processor and 8GB RAM, running the Windows operating system. The software environment includes Node.js v22.14.0, Electron 138 (headless) browser, Cypress version 15.8.2 test framework, Conduit RealWorld open-source web application under test, and Headless browser execution mode. Before each execution, the application state and test data were reset to a known baseline to ensure consistency.

### B. *Data Processing Pipeline*

In this paper, we adopt an empirical, evaluation-oriented case- study process to assess the effectiveness of automated E2E testing using Cypress. Case study is "an empirical method aimed at investigating contemporary phenomena in their context" **[17]**. A case study research method is selected because the phenomenon under investigation (the E2E automated testing) is tightly coupled with the context in which it is applied. In real-world software system, test automation behaviour cannot be isolated from user workflows, application structures and executed environment. The aim of this work is to evaluate the performance of an existing automation solution within a realistic software development context. The process is demonstrated in Fig. 6.

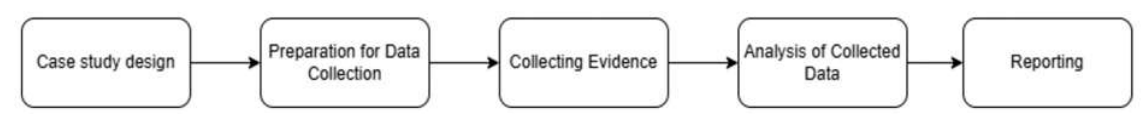

Fig. 6. Data Process Pipeline

The first step is to plan the case study objectives of what are expected to be achieved in the case study, build research questions, and choose metrics. After that, procedures and protocols for data collection are defined. The third step includes executing the test and collecting data, then the data is analyzed in step 5. The final step is reporting the result and writing conclusion. All data are recorded in Google Sheet.

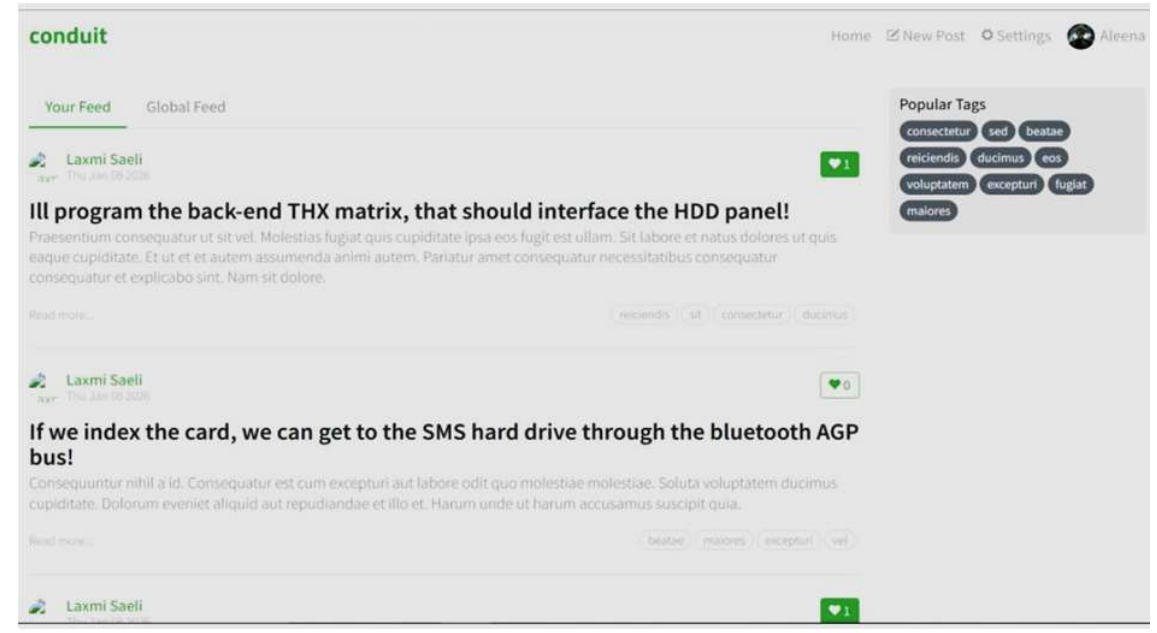

Fig. 7. Homepage of Conduit RealWorld

### C. *Web application*

The case study is conducted on Conduit RealWorld, which is a medium-sized social blogging platform (i.e. a Medium.com clone) designed to demonstrate typical

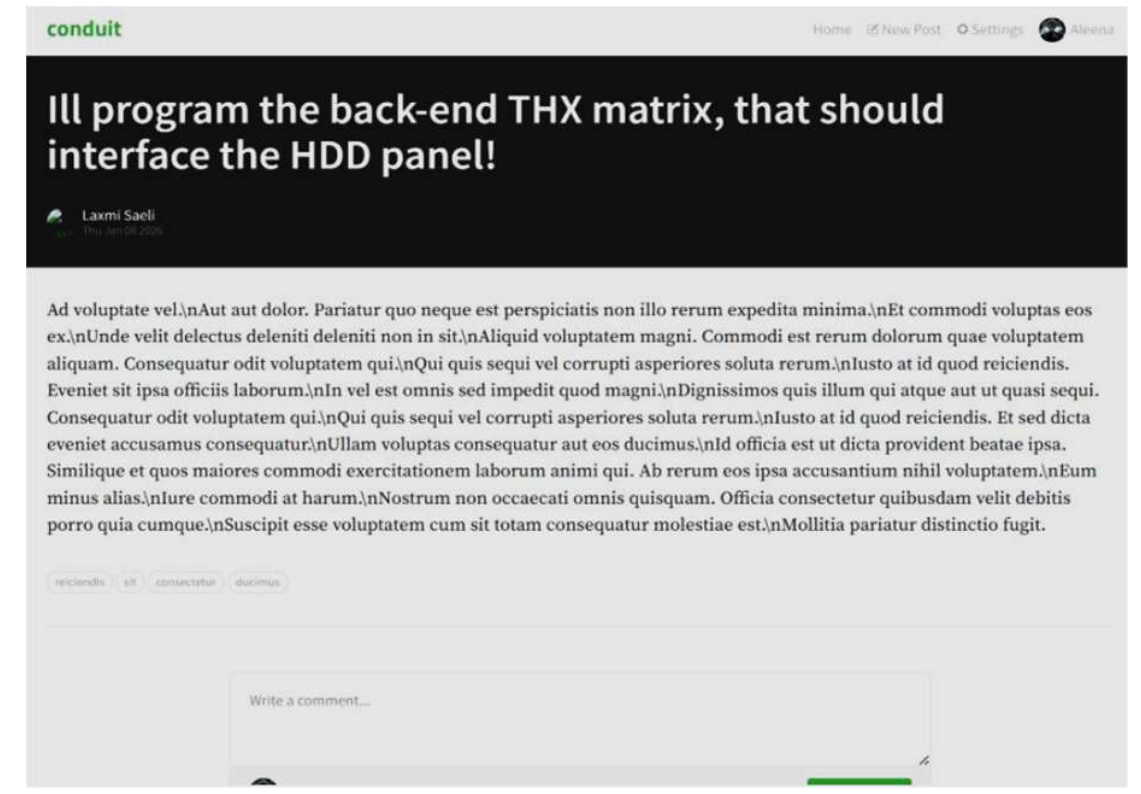

Fig. 8. Article detail page

architecture and functional patterns in modern web applications. The list of general features include:

- Authenticate users via JWT
- User state and session
- CRUD Articles
- CR-D Comments on articles
- List of articles pagination
- Favorite articles
- Follow other users

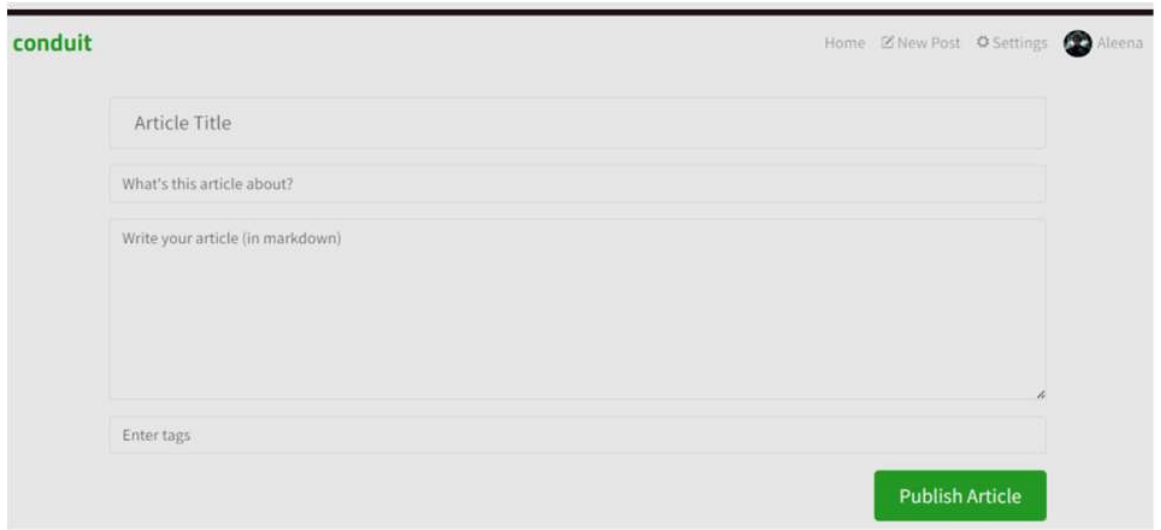

Fig. 9. Article editor page

Fig. 7 displays the main Homepage of the application with User Feed and Global Feed of articles. Fig. 8 displays a typical article detail page where user can view article content and perform comment action. Fig. 9 displays the article edit page where users can create a new article or edit their article. Fig. 10 displays the user settings where users can edit their personal information. This open-source web application is selected due to being a full-stack application with relatively complex UI and workflows, providing an excellent environment for E2E testing research.

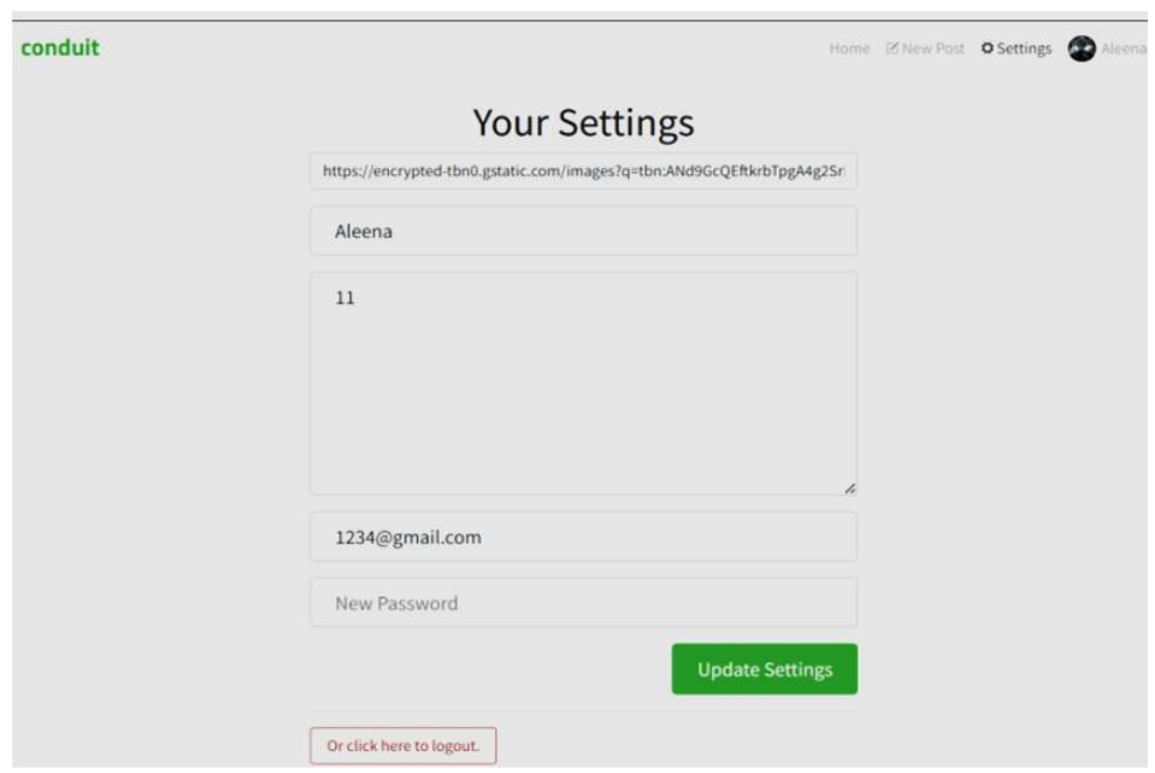

Fig. 10. Settings page

### D. *Test cases*

The E2E test suite consists of 27 test cases covering the core functional workflows of the Conduit platform. These cases are distributed across six main features: Authentication (Sign up/Sign in), Article CRUD operations, Favorite/Unfavorite articles, Follow/Unfollow authors, Comment management (Create/Read/Delete), and User Settings update. The list of 27 test cases is described in Appendix A.

### E. *Evaluation metrics*

There are 3 requirements for performance measurement when we run the experiments. That is,

***RQ1.*** *To what extent does Cypress-based end-to-end automated testing achieve execution velocity that supports iterative development in an open-source web application?*

For RQ1, we use formulas 1 and 2 to reason for execution velocity.

*1) Average Execution Time:* is average value of the total test run time across n executions. This index helps eliminate random variables and establish a stable representative number for each method.

$$\bar{T} = \frac{\sum_{i=1}^{n}(t_{end,i} - t_{start,i})}{n} \quad (1)$$

Where $(t_{end,i} - t_{start,i})$ is the execution time of the $i^{th}$ run, $n$ is the total number of experimental runs.

*2) Time reduction percentage.* measures the effectiveness of automated testing compared to manual testing, expressed as the percentage of time saved per iteration.

$$\Delta T = \left(\frac{\bar{T}_{manua} - \bar{T}_{auto}}{\bar{T}_{manual}}\right) \times 100\% \quad (2)$$

where $\bar{T}_{manual}$ is average execution time of manual testing, $\bar{T}_{auto}$ is average execution time of automated testing.

***RQ2.*** *How reliable and stable is Cypress-based automated testing across repeated executions, particularly in terms of test flakiness and false failure rates?*

For RQ2, we use formula 3 to reason the reliability.

*3) Flakiness frequency.* measures the frequency of discrete test failures that are not caused by changes in the application under test. A test case is considered flaky if it passes in one execution and then fails in another execution under the same condition.

$$F_{flaky} = \left(\frac{number\ of\ flaky\ cases}{total\ number\ of\ cases}\right) \times 100\% \quad (3)$$

*RQ3. How does the use of best practice element locating strategies affect the maintainability of Cypress test suites when the application interface changes?*

For RQ3, we use formula 4 to infer the maintainability.

*4) Time to repair.* is the time required to restore a failing automated test suite to a passing state after an application change. The time is measured using wall-clock time, starting from the moment a test failure is first observed after an application change, until the test suite passes again after the necessary test code modifications.

$$E = T_{pass} - T_{fail} \quad (4)$$

In addition, to evaluate execution time and stability metrics in RQ1 and RQ2, the test suite is executed repeatedly 20 times under the same condition. This repetition is designed to account for environmental variability and to identify flakiness. By executing 20 times, our work can calculate a stable mean execution time and observe the consistency of the pass rate. Therefore, the results are not influenced by isolated performance spikes or temporary system latencies. Manual testing for RQ1 is executed 3 times to mitigate potential measurement outliers (such as manual stopwatch delays or minor human distractions) while avoiding the confounding effects of human fatigue which would arise from excessive repetition. Application state is reset to a known baseline before each execution, the full Cypress suite is executed headlessly in the terminal, and the result of each execution is recorded automatically by the Cypress test runner. Total execution time, pass/fail status and individual outcomes of each test are recorded.

## V. Result Findings

### *A. Evaluating the Execution Speed for Iterative Development (RQ1)*

To answer the question of the iterative development capabilities of automated test suites, our work recorded the execution time of 27 E2E scenarios on the Cypress platform over 20 consecutive runs and compared it to the standard manual execution time. The results are presented in Table I.

Experimental results showed that the completion time of the automated test suite fluctuated steadily between 1 minute 20 seconds and 1 minute 39 seconds, with an average value of 1 minute 28 seconds (88 seconds). Meanwhile, the average time for manually executing the same scenario was 6 minutes 8 seconds (368 seconds) (see Table II). According to the formula, the execution time reduction rate is

$$\Delta T = \left(\frac{368s - 88s}{368s}\right) \times 100\% \approx 76.08\%$$

TABLE I. Results of 20 executive runs of 27 test cases

| **Execution ID** | **Passing rate** | **Time of execution (min)** | **Case failure ID** | **Fail feature** |
|---|---|---|---|---|
| [E-1] | 97% | 1:20 | #17 | Favorite Article |
| [E-2] | 100% | 1:24 | - | - |
| [E-3] | 100% | 1:27 | - | - |
| [E-4] | 97% | 1:29 | #18 | Favorite Article |
| [E-5] | 100% | 1:35 | - | - |
| [E-6] | 93% | 1:28 | #17, #18 | Favorite Article |
| [E-7] | 100% | 1:26 | - | - |
| [E-8] | 97% | 1:23 | #17 | Favorite Article |
| [E-9] | 93% | 1:30 | #17, #18 | Favorite Article |
| [E-10] | 93% | 1:34 | #17, #18 | Favorite Article |
| [E-11] | 97% | 1:39 | #18 | Favorite Article |
| [E-12] | 100% | 1:28 | - | - |
| [E-13] | 100% | 1:27 | - | - |
| [E-14] | 97% | 1:23 | #17 | Favorite Article |
| [E-15] | 97% | 1:25 | #17 | Favorite Article |
| [E-16] | 93% | 1:25 | #17, #18 | Favorite Article |
| [E-17] | 97% | 1:31 | #17 | Favorite Article |
| [E-18] | 93% | 1:38 | #17, #18 | Favorite Article |
| [E-19] | 97% | 1:29 | #18 | Favorite Article |
| [E-20] | 97% | 1:26 | #18 | Favorite Articl |

When comparing Cypress's test time (averaging 1 minute 28 seconds) to manual execution time (6 minutes 8 seconds), we can see a reduction in execution time of approximately 76.08%. The results indicate clearly the implementation of automation testing has made a great impact on the testing effort. In a workday where programmers are constantly updating code, and product is deployed in short development cycles, it makes a huge difference in workflows. With manual testing, we cannot afford to retest 27 different scenarios every time when a line of code is changed. Instead, we often have to wait for many changes to be accumulated before we test them all at once. This method inadvertently leads to error accumulation and makes it harder to find the root cause. However, with Cypress's speed, running tests after each commit becomes feasible and helps us enter the domain of continuous deployment in Agile's short cycle. Furthermore, when we do manual testing, the first testing might be very fast, but after that, the fatigue will reduce speed and accuracy. Therefore, using Cypress is not just about being faster but about freeing human effort. It saves time to more important tasks while the system is thoroughly tested.

TABLE II. Results of 3 times manual testing

| Execution ID | Passing rate | Time of execution | Case failure ID | Fail feature |
|---|---|---|---|---|
| [M-1] | 100% | 6:12 | - | - |
| [M-2] | 100% | 6:08 | - | - |
| [M-3] | 100% | 6:03 | - | - |

### *B. Evaluating the Reliability and Stability (RQ2)*

TABLE III. Comparison results between Suite 1 and Suite 2

| ID | Modification Type | Suite 1 | Suite 2 |
|---|---|---|---|
| CE1 | Text content modification | 129,5s | 0s |
| CE2 | Attribute value change | 126,78s | 0s |
| CE3 | CSS class refactoring | 132,6s | 0s |
| CE4 | DOM structure modification | 154,7s | 0s |
| CE5 | Element replacement | 159,4s | 0s |

To answer the question about the stability and reliability of Cypress-based automated testing across repeated executions, our work recorded the total times of failures and detail of which cases failed (see Table I). Based on the recorded data, the undetermined error rate of the entire system was calculated as follows:

$$F_{flaky} = \left(\frac{number\ of\ flaky\ cases}{total\ number\ of\ cases}\right) \times 100\% = \left(\frac{2}{27}\right) \times 100\% \approx 7.4\%$$

Over 20 consecutive cycles, the Cypress test suite demonstrated that while 92.6% of the scenarios remained stable, the Favorite Article function (test cases #17 and #18) showed fluctuating results which were sometimes passing and failing without any pattern. Thus, the failures are likely influenced by timing or shared state dependencies. Therefore, they are common causes of flakiness in end-to-end testing. As the Favorite Article functionality involves backend update operations and asynchronous UI rendering, the observed instability is acceptable and reflects the complexity of the functionality. It also stemmed from limitations of the testing framework. Therefore, the overall results show that Cypress-based automated testing performs stably for most functionalities when executed repeatedly. At the same time, the occurrence of a few flaky tests underscores the importance of careful test design and attention to stability, especially when functionalities with complex interactions are tested, as test instability can undermine the core value of automated testing **[19]**.

### C. *Evaluating the Maintainability (RQ3)*

To answer question about how the use of best-practice element locating strategies affects the maintainability of Cypress test suites, our work used two test suites with the same testing logic but different element locating strategies. The first test suite uses traditional locators based on text content, HTML attributes, CSS classes, and DOM structure **[18]**. The second test suite uses a robust locator based on the data-cy attribute. The table III presents the data collected from the experiments.

The results show that test suites using traditional locators are affected by all types of interface changes surveyed **[18]**. Even small changes, such as changes to text content or attribute values, cause tests to fail and require manual correction. In contrast, utilizing the robust locator is isolated from
the user interface updates, and the test suite remains unaffected by any interface change. The fact that these locators are not dependent on the structure or presentation of the interface, which helps the test suite to continue functioning normally without modification. Although the repair time per case for traditional test suites only takes a few minutes due to the organization of test structure in Page Object Model, this result shows that maintenance costs can accumulate significantly when the number of test cases is large or when the interface changes frequently. Therefore, it is particularly important in the context of E2E testing, as tests frequently cover lengthy business flows and rely largely on user interfaces. The robust locators significantly reduce the effort required to maintain the test suite over the long term, even when the application continues to evolve and change.

### D. *Threats to Validity*

There are several threats to the validity of this study. First, the evaluation was conducted on a single open-source web application (Conduit RealWorld), which limits the generalizability of the results to applications of different sizes or architectures. Second, the flakiness observed in end-to-end testing can be influenced by external factors such as network latency or backend state changes that cannot be completely isolated. Finally, the evaluation of maintainability based on repair time may be subject to the tester's experience bias.

## VI. CONCLUSION

Our work is to evaluate the effectiveness of end-to-end (E2E) automated testing on an open-source web application, through a case study using the Cypress framework. The effectiveness of automated testing is considered in three main aspects: execution time, stability, and maintainability of the test suite. Nevertheless, the effectiveness is heavily dependent on the test suite's design and maintenance, automated testing cannot totally eliminate concerns such as flakiness or maintenance expenses. However, when it is done correctly, it can add significant value to the software development process.

For the future works, we will examine projects with different sizes, domains, and architectural styles in order to improve the generalizability of the results. We will incorporate backend log analysis, network delay evaluation, or test execution order control to gain a deeper understanding of the factors contributing to flakiness in E2E testing. We plan to conduct experimental comparisons between Cypress and other E2E frameworks like Selenium or Playwright, to examine differences in execution efficiency, test stability, and maintenance effort across various development contexts.

APPENDIX A: 27 TEST CASES

| ID | Type | Description | Pre-condition | Steps | Expected Res |
|---|---|---|---|---|---|
| [TC-1] | Auth | User can register with valid credentials | User is not logged in.<br>User is on the Conduit home page.<br>The email address has not been registered previously. | 1.Navigate to the Sign up page.<br>2.Enter a valid username.<br>3.Enter a valid email address.<br>4.Enter a valid password.<br>5.Click the Sign up button. | Account is created; user is logged in |
| [TC-2] | Auth | Registration fails with existing email | An account with the email "123@gmail.com" already exists.<br>User is not logged in. | 1. Navigate to the Sign up page.<br>2. Enter a valid username.<br>3. Enter an email address that is already registered.<br>4. Enter a valid password.<br>5. Click the Sign up button. | Registration is rejected.<br>An error message indicating that the email is already in use is displayed.<br>User remains on the Sign up page. |
| [TC-3] | Auth | User can log in with valid credentials | A registered user account exists.<br>User is not logged in. | 1.Navigate to the Sign in page.<br>2.Enter a registered email address.<br>3.Enter the correct password.<br>4.Click the Sign in button. | User is successfully logged in.<br>User is redirected to the home page.<br>The user's username appears in the navigation bar. |
| [TC-4] | Auth | Login fails with invalid credentials | A registered user account exists.<br>User is not logged in. | 1. Navigate to the Sign in page.<br>2. Enter a registered email address.<br>3. Enter an incorrect password.<br>4. Click the Sign in button. | Login attempt fails.<br>An error message indicating invalid credentials is displayed.<br>User remains on the Sign in page. |
| [TC-5] | Auth | User cannot login with empty email | User is not logged in. | 1. Navigate to the Sign in page.<br>2. Leave email field blank.<br>3. Enter an incorrect password.<br>4. Click the Sign in button. | Login attempt fails.<br>An error message indicating email cannot be blanked is displayed.<br>User remains on the Sign in page. |
| [TC-6] | Auth | User cannot login with empty password | User is not logged in. | 1. Navigate to the Sign in page.<br>2. Enter a valid email address.<br>3. Leave password field empty.<br>4. Click the Sign in button. | Login attempt fails.<br>An error message indicating password cannot be blanked is displayed.<br>User remains on the Sign in page. |
| [TC-7] | Auth | User can log out successfully | User is logged in. | 1. Click on the Settings link in the navigation bar.<br>2. Click the Logout button. | User is logged out.<br>User is redirected to the home page.<br>Sign in and Sign up links are visible in the navigation bar. |
| [TC-8] | Auth | Session persists after page reload | User is logged in. | Reload the current page using the browser refresh function. | User remains logged in.<br>User's username is still displayed in the navigation bar.<br>No redirection to the login page occurs. |
| [TC-9] | Article CRUD | Logged-in user can create an article | User is logged in. | 1. Click the New Article button.<br>2. Enter a valid article title.<br>3. Enter a valid article description.<br>4. Enter article content in the body field.<br>5. Click the Publish Article button. | Article is successfully created.<br>User is redirected to the article detail page.<br>The newly created article is displayed with correct content. |
| [TC-10] | Article CRUD | Article creation fails with missing all required fields | User is logged in. | 1. Click the New Article button.<br>2. Leave all of required fields empty (e.g., title or body).<br>3. Click the Publish Article button. | Validation error message "title can't be blank" is displayed.<br>User remains on the editor page instead of navigating to article page. |
| [TC-11] | Article CRUD | Article creation fails with missing description field | User is logged in. | 1. Click the New Article button.<br>2. Leave desciprtion field empty<br>3. Fill other fields with valid credential<br>3. Click the Publish Article button. | Validation error message "title can't be blank" is displayed.<br>User remains on the editor page instead of navigating to article page. |
| [TC-12] | Article CRUD | Article creation fails with missing body field | User is logged in. | 1. Click the New Article button.<br>2. Leave body field empty<br>3. Fill other fields with valid credential<br>3. Click the Publish Article button. | Validation error message "title can't be blank" is displayed.<br>User remains on the editor page instead of navigating to article page. |
| [TC-13] | Article CRUD | Article creation success with tag field empty | User is logged in. | 1. Click the New Article button.<br>2. Leave tag field empty<br>3. Fill other fields with valid credential<br>3. Click the Publish Article button. | Article is successfully created.<br>User is redirected to the article detail page.<br>The newly created article is displayed with correct content. |
| [TC-14] | Article CRUD | Author can edit own article | User is logged in.<br>User is the author of an existing article. | 1. Navigate to the article detail page of the user's own article.<br>2. Click the Edit Article button.<br>3. Modify the article content. | Article is successfully updated.<br>Updated content is displayed on the article detail page. |
| [TC-15] | Article CRUD | Author can delete own article | User A is logged in.<br>User A has at least one article published. | 1.Go to User A's article list or directly to the article detail page.<br>2.Click the Delete Article button.<br>3.Confirm deletion if a prompt appears. | Article is removed from the feed/profile.<br>User is redirected to the home or article list page.<br>The article does not appear in My articles list in User's profile page |
| [TC-16] | Permissions - A | Unauthorized user cannot create article | User is not logged in (guest). | 1.Navigate to Create New Article page.<br>2.Attempt to submit a new article with title, description, and body. | User is shown an authorization error.<br>Article is not created in the backend. |
| [TC-17] | Favorite/Unfavo | User can favorite an article | User is logged in.<br>At least one article exists that user did not favorite. | 1.Navigate to the article list.<br>2.Click the Favorite (heart) button. | Heart button changes state (filled or active).<br>Favorites count increments by 1.<br>Article appears on user's favorites page |
| [TC-18] | Favorite/Unfavo | User can unfavorite an article | User is logged in.<br>Article is already favorited by the user. | 1.Navigate to the favorited article.<br>2.Click the Unfavorite (heart) button. | Heart button reverts to inactive/unfilled.<br>Favorites count decrements by 1.<br>Article no longer appears in the Favorites tab |
| [TC-19] | Follow/Unfollow | User can follow article author | User is logged in.<br>Article author is not yet followed by the user. | 1.Navigate to the author's profile via article.<br>2.Click Follow button. | Button changes to "Unfollow".<br>Author appears in user's following list. |
| [TC-20] | Follow/Unfollow | User can unfollow article author | User is logged in.<br>User already follows the author. | 1.Navigate to the author's profile.<br>2.Click Unfollow button. | Button changes to "Follow".<br>Author no longer appears in user's following list. |
| [TC-21] | Follow/Unfollow | Logout user (Guest) cannot follow author | User is logged out.<br>At least 1 seed author exists in global feed. | 1Navigate to author's profile page.<br>2.Attempt to follow by clicking "Follow"<br>3.Observe the UI response<br>4.Inspect the Network tab for any API requests triggered by the action. | The UI does not reflect a successful follow state. Button does not reflects a followed state<br>API /follow returns 401 unauthorized |
| [TC-22] | Comment | Logged-in user can post a comment | User A is logged in<br>An article exists and is accessible<br>Comment input field is | 1.Navigate to an article detail page<br>2.Enter a valid comment in the comment input field<br>3.Click the Post Comment button | The comment is successfully submitted<br>The new comment appears in the comment list |
| [TC-23] | Comment | Logged-in user cannot post an empty comment | User A is logged in<br>An article exists and is accessible<br>Comment input field is visible on the article page | 1.Navigate to an article detail page<br>2.Leave the comment box empty<br>Click the Post Comment button | No empty comment is submitted<br>Comment section does not include empty comment |
| [TC-24] | Comment | Comment author can delete own comment | User A is logged in<br>User A has previously posted a comment on an article<br>The comment is visible on the article page | 1.Navigate to the article containing User A's comment<br>2.Locate User A's comment<br>3.Click the Delete (trash) button on the comment | The comment is removed from the comment list<br>The comment does not reappear after page refresh |
| [TC-25] | Comment | Logged-out user cannot add comment | User is logged out<br>An article exists and is accessible | 1.Navigate to the article detail page<br>2.Attempt to enter a comment<br>3.Attempt to submit the comment | No comment input is available<br>No comment is added to the list |
| [TC-26] | Settings | User can update bio | User is logged in<br>User profile settings page is accessible | 1.Navigate to the Settings page<br>2.Modify the Bio field with valid text<br>3.Click Update Settings<br>4.Navigate to the user's profile page | Bio is updated successfully<br>User is redirected to Homepage<br>Updated bio is displayed on the profile page<br>Bio persists after page refresh or re-login |
| [TC-27] | Settings | User can update profile image URL | User is logged in<br>Settings page is accessible<br>A valid image URL is available | 1.Navigate to the Settings page<br>2.Enter a valid URL in the Profile Image URL field<br>3.Click Update Settings<br>4.Navigate to the user's profile page | Profile image is updated successfully<br>User is redirected to Homepage<br>The new image is displayed on the profile page<br>Updated image persists after refresh |